\documentclass[epj]{svjour}
\usepackage[T1]{fontenc}
\usepackage{amsmath}%
\usepackage{amssymb}%
\usepackage{amsfonts}%
\usepackage{mathtools}%
\usepackage{mathrsfs}%
\usepackage{graphicx}%
\usepackage[caption=false, singlelinecheck=false,
justification=raggedright]{subfig}%
\usepackage{xspace}%
\usepackage{txfonts}%
\usepackage{soul}%
\usepackage{paralist}%
\usepackage{nicefrac}%
\usepackage{cite}%
\usepackage[colorlinks, urlcolor=black, citecolor=black, linkcolor=black]{hyperref}%
\newcommand{\modulus}[1]{\lvert #1 \rvert}
\newcommand{\Br}{\ensuremath{\textrm{Br}}}
\newcommand{\overbar}[1]{\kern 0.2em\overline{\kern -0.2em #1}{}\xspace}
\newcommand*{\DOI}[2]{\href{http://dx.doi.org/\detokenize{#1}}{#2}}
\begin{document}
\title{Probing sterile neutrinos in $B$ ($D$) meson decays at Belle~II (BESIII)}

\author{C. S. Kim\inst{1,2}\thanks{Email at: cskim@yonsei.ac.kr}, Youngjoon Kwon\inst{1}\thanks{Email at: yjkwon63@yonsei.ac.kr}, Donghun Lee\inst{1}\thanks{Email at: donghun.lee@yonsei.ac.kr}, Sechul
Oh\inst{3}\thanks{Email at: scohph@yonsei.ac.kr} \and Dibyakrupa Sahoo\inst{1}\thanks{Email at: sahoodibya@yonsei.ac.kr}}%
\authorrunning{Kim, Kwon, Lee, Oh and Sahoo}

\institute{Department of Physics and IPAP, Yonsei University, Seoul 03722, Korea 
\and Institute of High Energy Physics, Dongshin University, Naju 58245, Korea
\and University College, Yonsei University, Incheon 21983, Korea}%

\date{Received: date / Revised version: date}%
%
\abstract{We present, how a systematic study of $B \to D\ell N$ ($D \to K \ell
N$) decays with $\ell=\mu,\tau$, at Belle~II (BESIII) can provide unambiguous
signature of a heavy neutrino $N$ and/or constrain its mixing with active
neutrinos $\nu_\ell$, which is parameterized by $\left| U_{\ell N} \right|^2$.
Our constraint on $\lvert U_{\mu N} \rvert^2$ that can be achieved from the full
Belle II data is comparable with what can be obtained from the much larger data
set of the upgraded LHCb. Additionally, our method offers better constraint on
$\lvert U_{\mu N} \rvert^2$ for mass of sterile neutrino $m_N < 2$~GeV. We can
also probe the Dirac and Majorana nature of $N$ by observing the sequential
decay of $N$, including suppression from observation of a displaced vertex as
well as helicity flip, for Majorana $N$.%
} 
\maketitle %

\section{Introduction}\label{sec:intro}

In many new physics theories, called see-saw mechanisms, there are one or more
heavier cousins of the active flavor neutrinos $\nu_\ell$ ($\ell=e,\mu,\tau$)
which do not have any interaction with Standard Model (SM) particles except
mixing with the active neutrinos. These heavy neutrinos are named as sterile
neutrinos which can be either Dirac or Majorana fermions. Among the varieties of
new physics scenarios where sterile neutrinos appear, the original seesaw
mechanism~\cite{seesaw} predicts their mass to be much larger than $1$~TeV. In
other seesaw models heavy neutrinos can have mass in a very large range, from
$\sim 0.1$~TeV to $1$~TeV~\cite{1TeVNu}, or close to about
$1$~GeV~\cite{1GeVNu}, or even at keV scale \cite{Dodelson:1993je} or eV scale
\cite{Palazzo:2013me}. The mixing parameters, $U_{\ell N}$, which describe the
strength of mixing between a sterile (heavy) neutrino $N$ with the SM flavor
(light) neutrinos $\nu_{\ell}$ are constrained by various experimental data
depending on the mass of $N$ (see Refs.~\cite{Deppisch:2015qwa,Cvetic:2019shl}
for further details and references).

The neutrinos are the only fermions which can be their own anti-particles, i.e.\
behave as Majorana fermions. Ascertaining their Dirac or Majorana nature is one
of the most important issues in neutrino physics. It is well known that Dirac
neutrinos can participate only in the lepton number conserving (LNC) processes,
while Majorana neutrinos can get involved in both lepton number violating (LNV)
and LNC processes. Therefore, to investigate the Majorana nature of neutrinos,
many attempts have been made at studying various LNV processes, including
neutrinoless double beta decay ($0 \nu \beta \beta$)~\cite{nndb1}, specific LNV
processes at LHC~\cite{scatt1,scatt2,scatt3,KimLHC}, LNV $\tau$ lepton
decays~\cite{tau}, and LNV rare meson
decays~\cite{RMDs,HKS,Atre,CDKK,Mand,BtoDlN,B2DlN}, e.g.\ rare LNV decays of
$K$, $D_{(s)}$, $B_{(c)}$ mesons have been studied extensively in literature. In
particular, semileptonic decays such as $B \to D \ell \ell \pi$ and $B \to \ell
\ell \pi$ were explored in Refs.~\cite{BtoDlN,B2DlN} to not only distinguish
between Dirac and Majorana signatures, but also constrain $\modulus{U_{\ell
N}}^2$, without considering, in detail, the feasibility of observation of these
decays inside a sizable detector including the helicity flip for Majorana case.

Our main significant result in this paper is the stringent constraint that can
be put on $\modulus{U_{\ell N}}^2$, especially on $\modulus{U_{\mu N}}^2$, from
non-observation of the decays $B \to D \ell N$, without considering the
sequential decay of $N$. This simple strategy has, however, remained unexplored
in the currently existing literature. Instead of considering two-body leptonic
decays $B^+ \to \ell^+ N$, similar to existing studies on $\pi^+ \left(\text{or
} K^+ \right) \to \ell^+ N$ which look for mono-energetic $\ell^+$ to constrain
$\modulus{U_{\ell N}}^2$ \cite{peak-search}, we have considered the three-body
semileptonic decays $B \to D \ell N$ which have bigger branching ratios in a
larger mass range. The reach of our study to constrain $\modulus{U_{\mu N}}^2$
and $\modulus{U_{\tau N}}^2$ is better by an order of magnitude from existing
experimental constraints in certain mass ranges of interest. Interestingly, our
constraint on $\modulus{U_{\mu N}}^2$ obtained by considering only $\sim 4.8
\times 10^8$ events of fully reconstructed $B\to D\mu N$ decays at Belle II
\cite{Kou:2018nap} is comparable with the constraint achievable from $4.8 \times
10^{12}$ events of $B\to D \mu\mu\pi$ decays at upgraded LHCb
\cite{Cvetic:2019shl}. Although the missing sterile neutrino search gives the
stringent constraint on $\modulus{U_{\ell N}}^2$, it can not distinguish Dirac
and Majorana neutrinos. Therefore, we also study the sequential decay of $N$
with a displaced vertex signature for probing its Majorana nature, and consider
the important but otherwise overlooked effect of helicity flip for sterile
neutrinos. Despite the suppression coming from observation of displaced vertices
as well as the helicity flip, we find that heavier and less energetic neutrinos
have a bigger chance of decaying inside a detector with decay length $\leqslant
1$~m, provided they exist. Finally, we present an estimate of $\modulus{U_{\mu
N}}^2$ in the case of observation of LNC $B \to D \mu^+ \mu^- \pi^+$ decay in
Belle~II. The LNV mode $B \to D \mu^+ \mu^+ \pi^-$ receives additional
suppression from helicity flip.

This paper is organized as follows. In Sec.~\ref{sec:choice} we provide the
logical basis for considering the decays $B \to D \ell N$ or $D \to K \ell N$
and the advantages they offer over choosing any other processes. In
Sec.~\ref{sec:constraint} we show how well the mixing parameters
$\modulus{U_{\mu N}}^2$ and $\modulus{U_{\tau N}}^2$ can be probed by using
these decays at Belle~II. We also provide a short discussion, via example, on
possible SM background processes and how they can be distinguished from the
signal events. This is followed by a discussion in Sec.~\ref{sec:majorana} on
how the Majorana nature of neutrino could be probed, and whether it is possible
to do such a study. Finally we conclude in Sec.~\ref{sec:conclusion}
highlighting the important features of our paper.

\section{Choosing appropriate production modes}\label{sec:choice}

We aim to find a process that would
\begin{inparaenum}[(1)]
\item unambiguously probe the presence of sterile neutrino $N$, free from
any other new physics possibilities, and %
\item constrain the mixing of $N$ with active neutrinos.
\end{inparaenum}
In this regard, we find it helpful to keep in mind the following four cardinal
aspects of sterile neutrino. Any candidate for sterile neutrino would have
\begin{inparaenum}[(i)]
\item electric charge $=0$, %
\item spin $=\nicefrac{1}{2}$, %
\item mass $\gg 0$, and %
\item possibly long life-time.
\end{inparaenum}
The fact that the sterile neutrino would most likely remain undetected at its
point of production, is in fact the experimental consequence of its possibly
long life-time and electrically neutral nature. This manifests as `missing
momentum' in any process which would have sterile neutrino(s) in the final
state, just like the case with ordinary active neutrino(s) in any final state.

As examples of processes with `missing momentum' we can consider meson decays
such as $B \to D^{(*)} \mu X$, $D \to K^{(*)}\mu X$, $B \to K X$ etc., where $X$
denotes missing (i.e.\ undetected) particle(s) other than active neutrinos. A
simple analysis of spin would suffice to illustrate the fact that for decays
such as $B \to D^* \mu X$ and $D \to K^* \mu X$, the spin of $X$ is ambiguous:
it could be $\nicefrac{1}{2}$ or $\nicefrac{3}{2}$. Similarly, in the decay $B
\to K X$, the `missing' $X$ would indeed be made of (at least) two invisible
particles and their individual spins could be $0$, $\nicefrac{1}{2}$ or
$\nicefrac{3}{2}$. In order to avoid such ambiguities, we shall refrain from
considering these types of decays in this paper. Thus we are left to consider
decays of the type $B \to D \mu N$ and $D \to K \mu N$ as decay modes suitable
for discovery of sterile neutrino $N$.

It should be noted that the literature dealing with sterile heavy neutrino
searches is replete with LNC and LNV processes mediated by sterile neutrino $N$,
such as $B \to D \ell \ell \pi$, $B \to \ell \ell \pi$, $\tau \to \pi \ell_1
\ell_2 \nu$ etc.\ where $\ell, \ell_{1,2}=e,\mu$. These processes, can also take
place via other new physics possibilities, such as exotic scalars, vectors or
lepto-quarks. Nevertheless, the LNV processes mediated by $N$ constitute the
only known reliable methodology to probe the conjectured Majorana nature of $N$.
Notwithstanding the importance, these neutrino mediated decays are suppressed
from associated displaced vertices, branching ratio of sequential decay of $N$
as well as helicity flip of $N$ (relevant in case of LNV modes only). Therefore,
when concerned with the discovery prospect of sterile neutrino, we shall refrain
from considering the sequential decay of $N$. Once an unambiguous signature of
existence of sterile neutrino $N$ is obtained, study of its Dirac or Majorana
nature becomes highly relevant, and in this context we would consider the
feasibility of the complete LNC and LNV modes which take into account the
sequential decay of $N$.

Let us now analyze the decays $B \to D \mu N$ (or $D \to K \mu N$) with $N$
remaining undetected in the detector. We are interested in the scenario where
there is one light sterile neutrino with mass $m_N \leqslant 3.3$~GeV. The
criteria of massive sterile neutrino helps us to eliminate background events to
our processes. The decay $B \to D \mu N$, for example, can receive background
from the decay $B \to D \mu \nu_{\mu} \nu \overbar{\nu}$ or $B \to D^* (\to D
\pi_\textrm{soft}/\gamma_\textrm{soft}) \mu \nu$, where the soft pion
($\pi_\textrm{soft}$) or soft photon ($\gamma_\textrm{soft}$) arising from
sequential decay of $D^*$ are not detected by the detector. However, the
invariant mass obtained from the missing 4-momenta of background process would
vary significantly from one event to another unlike the signal case which would
be centered about the fixed mass of $N$.

In order to obtain the `missing mass' ($m_\textrm{miss}$) in the processes, say
$B \to D \mu + \textrm{`missing'}$ that includes the signal events for $B \to D
\mu N$ also, we need to know the 4-momenta of initial $B$ meson ($p_B$), final
$D$ meson ($p_D$) and $\mu$ ($p_\mu$):
\begin{equation}
m_\textrm{miss} \equiv
\sqrt{p_\textrm{miss}^2} = \sqrt{\left(p_B - p_D - p_\mu \right)^2}.
\end{equation}
For signal events only $m_\textrm{miss} = m_N$. This methodology is applicable
in experiments such as Belle~II or BESIII where $B$ and $D$ mesons are pair
produced along with $\overbar{B}$ and $\overbar{D}$ from the decays
$\Upsilon(4S) \to B\overbar{B}$ and $\psi(3770) \to D\overbar{D}$ respectively,
and the 4-momentum of $\overbar{B}$, $\overbar{D}$ can be precisely measured by
full hadronic reconstruction. It should be, therefore, clear that our
methodology is not applicable for experiments such as LHCb where the initial
4-momentum of the $B$ or $D$ meson can not be inferred without measuring the
4-momenta of the final particles arising from the $B$ or $D$ meson decay.
Furthermore, the minimum value of mass $m_N \neq 0$ that can be probed in our
approach is, therefore, constrained only by the experimental accuracy of
measurement of 4-momenta of $B$, $D$ and $\mu$. In the next section we would
provide a numerical comparison of a few observables (including $m_N$), for the
SM background decay $B \to D \mu \nu \pi_\textrm{soft}/\gamma_\textrm{soft}$ and
the signal decay $B \to D \mu N$, specifically in context of Belle~II. It is
important to note that $N$ may or may not decay inside a detector, depending on
its mass, energy and the size of the detector. If it decays inside the detector,
with noticeable displaced vertex, we can not only measure its 4-momentum
directly from its decay products, but also probe its Dirac or Majorana nature as
well as veto any background events for the decays under consideration.

Note that the decays $B^+ \to \tau^+ N$ and $B \to D \tau N$, where the
4-momentum of the tau lepton is reconstructed from its further sequential decay,
are less promising for our study, due to the presence of at least one
neutrino (or antineutrino) in the final state of all tau decays. Nevertheless,
taking into account that the tau 4-momentum could be measured accurately with a
smaller probability, we shall constrain $\modulus{U_{\tau N}}^2$ from $B \to D
\tau N$.

\section{Determining or constraining the value of $\modulus{U_{\mu N}}^2$ and $\modulus{U_{\tau N}}^2$}\label{sec:constraint}

The branching ratios of all the decay modes under
our consideration are directly proportional to the appropriate active-sterile
mixing parameter $\modulus{U_{\ell N}}^2$. We obtain the \textit{canonical
	branching ratio} of a decay, e.g.\ $B \to D \ell N$, by factoring out
$\modulus{U_{\ell N}}^2$ from the theoretically calculable branching
ratio~\cite{BtoDlN},
\begin{equation}
\underbar{\Br}\left(B\to D \ell N\right) = \frac{\Br\left(B\to D \ell N\right)}{\modulus{U_{\ell N}}^2}.
\end{equation}
Given the value of canonical branching ratio $\underbar{\Br}\left(B\to D \ell
N\right)$, the number of such decays observable in the detector ($N_{B \to D
\ell N}$) and the total number of fully reconstructed parent particles ($N_B$),
we can estimate $\modulus{U_{\ell N}}^2$ by
\begin{equation}
\modulus{U_{\ell N}}^2 = \frac{N_{B \to D \ell N}}{N_B \times \epsilon_D \times \epsilon_\ell \times
	\underbar{\Br}\left(B\to D\ell N\right)}~,
\end{equation}
where $\epsilon_D,\epsilon_\ell$ denote the efficiency to reconstruct the
$D,\ell$ in the signal side. In our numerical study discussed below we have
assumed $\epsilon_D = \epsilon_\mu = 1$, but $\epsilon_\tau = 0.001$ (the reason of which is given later).

\begin{figure}[ht]
\centering%
\subfloat[From the decay $B \to D \mu
N$.\label{fig:UmNSq}]{\includegraphics[width=0.9\linewidth, trim=0 1mm 2mm 2mm,
clip]{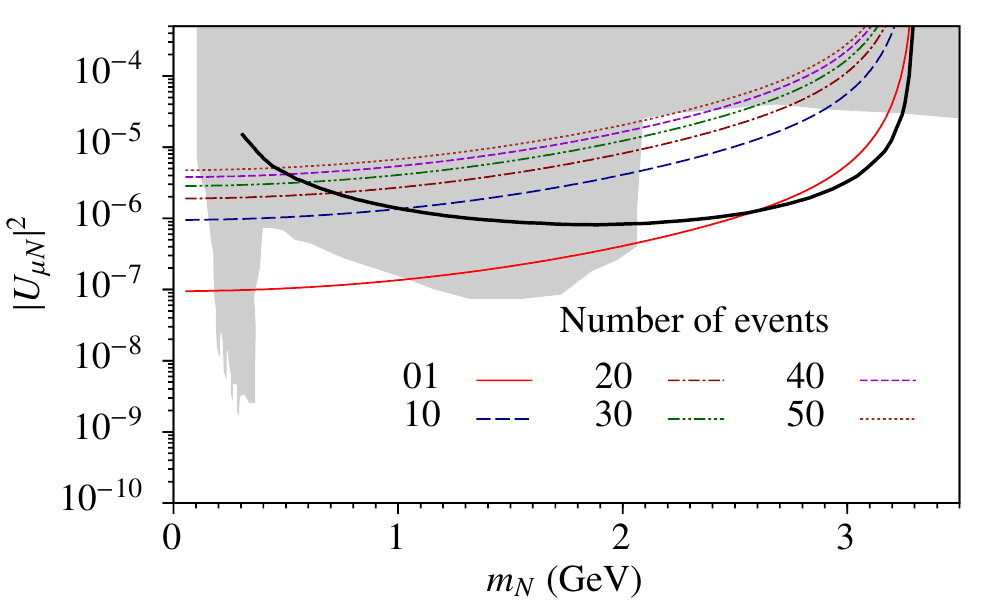}}\\%
\subfloat[From the decay $B \to D \tau
N$.\label{fig:UtNSq}]{\includegraphics[width=0.9\linewidth, trim=0 1mm 2mm 2mm,
clip]{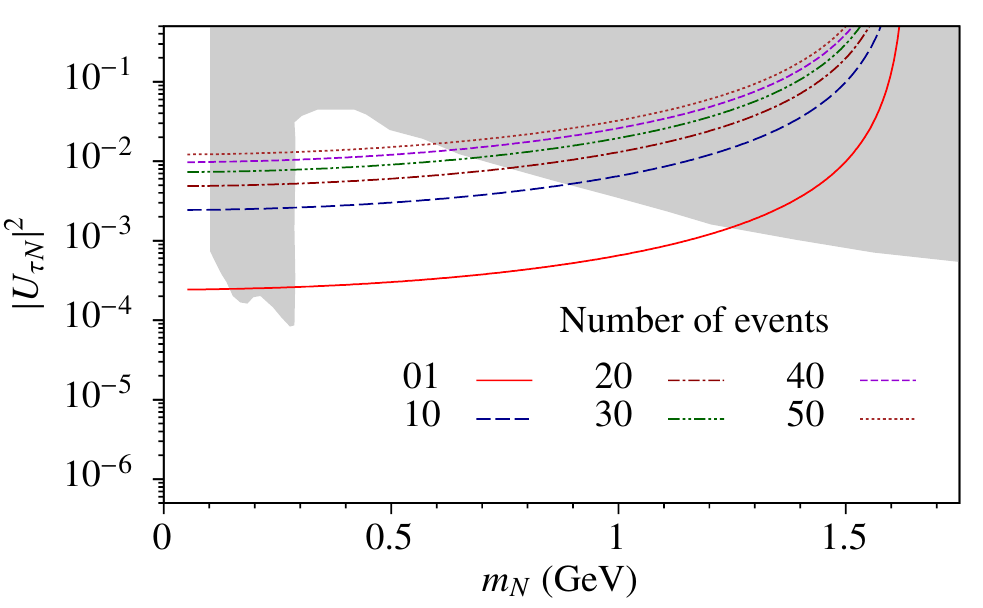}}%
\caption{Values of $\modulus{U_{\mu N}}^2$ and $\modulus{U_{\tau N}}^2$
\textit{estimated} from observed number of events (less than $50$ events) of the
decays $B \to D \mu N$ and $B \to D \tau N$, respectively, using the projected
number of $B$ decays at Belle~II and assuming $0.1\%$ chance of full reconstruction of $\tau$ from its decays. The $B \to D \ell N$ decays include both the charged and neutral modes. The thick solid line corresponds to the predicted $95\%$ C.L.\ upper-limit on $\modulus{U_{\mu N}}^2$~\cite{Cvetic:2019shl}, for $4.8 \times 10^{12}$ $B$ decay events at upgraded LHCb (with the decay $B \to D \mu \mu \pi$).}%
\label{fig:ULNSq-constraint}%
\end{figure}

For a numerical study we consider the decays $B \to D \mu N$ and $B \to D \tau
N$ in context of Belle~II experiment, which is poised to detect $10^{11}$ $B$
decay events~\cite{Kou:2018nap}. Out of these, about $0.61\%$ of charged $B$
events and $0.34\%$ of neutral $B$ events can be fully reconstructed from
hadronic tagging~\cite{Kou:2018nap}, so that only about $4.8 \times 10^8$ events
of $B$ decays get fully reconstructed. Considering only these $B$ decays, we are
able to estimate the value of $\modulus{U_{\mu N}}^2$, as shown in
Fig.~\ref{fig:UmNSq}, from possible observation of 50 events or less for $B \to
D \mu N$. It is easy to observe that in the mass range $\sim 2-3~\text{GeV}$ our
approach can provide stronger constraint, by about one order of magnitude, than
the existing experimental upper-limit (exclusion region at $\sim90\%$ C.L.\ from
various experiments is shown by the shaded region in gray). Fig.~\ref{fig:UmNSq}
also shows that our constraint is comparable with the $95\%$ C.L.\ upper-limit
on $\modulus{U_{\mu N}}^2$, shown as a thick solid line, predicted in
Ref.~\cite{Cvetic:2019shl} based on $4.8 \times 10^{12}$ $B$ decay events at
upgraded LHCb (with the decay $B \to D \mu \mu \pi$). For $m_N<2$~GeV (important
for light sterile neutrino searches) our constraint significantly surpasses the
above-mentioned constraint predicted for LHCb upgrade. This is primarily due to
the suppression factors affecting the observation of $B \to D \mu \mu \pi$
decays inside a finite-sized detector for smaller values of $m_N$ (see the next
section and Figs.~\ref{fig:NumDensity-FullEvents} and \ref{fig:feasibility}). It
is worth mentioning that the method proposed in this paper cannot be applied to
LHCb, for it requires full reconstruction of the rest of the event so that
missing energy-momentum can be used to extract the information about $N$.

Similarly, we can constrain $\modulus{U_{\tau N}}^2$ from number of observed $B
\to D \tau N$ decays if the 4-momentum of the final $\tau$ could be measured
accurately. In Fig.~\ref{fig:UtNSq}, we show estimations of $\modulus{U_{\tau
N}}^2$ as a function of $m_N$, for different values of observed $B \to D \tau N$
decay events. Here we have assumed that the 4-momenta of only $0.1\%$ of all the
$\tau$ decays could be precisely measured (e.g.\ more than 3-prong decays of
$\tau$ \cite{PDG2018}). It is clear from Fig.~\ref{fig:UtNSq} that our
constraint on $\modulus{U_{\tau N}}^2$ in the mass range $[0.3,1]$~GeV is more
stringent than the existing studies, by an order of magnitude in some $m_N$
region. It should be noted that if we can further improve the $\tau$
reconstruction efficiency, our current result would further improve.

\begin{figure*}[ht]
\centering%
\subfloat[Distribution of events with respect to $s$.
\label{fig:bg-s}]{\includegraphics[width=0.45\linewidth,keepaspectratio]{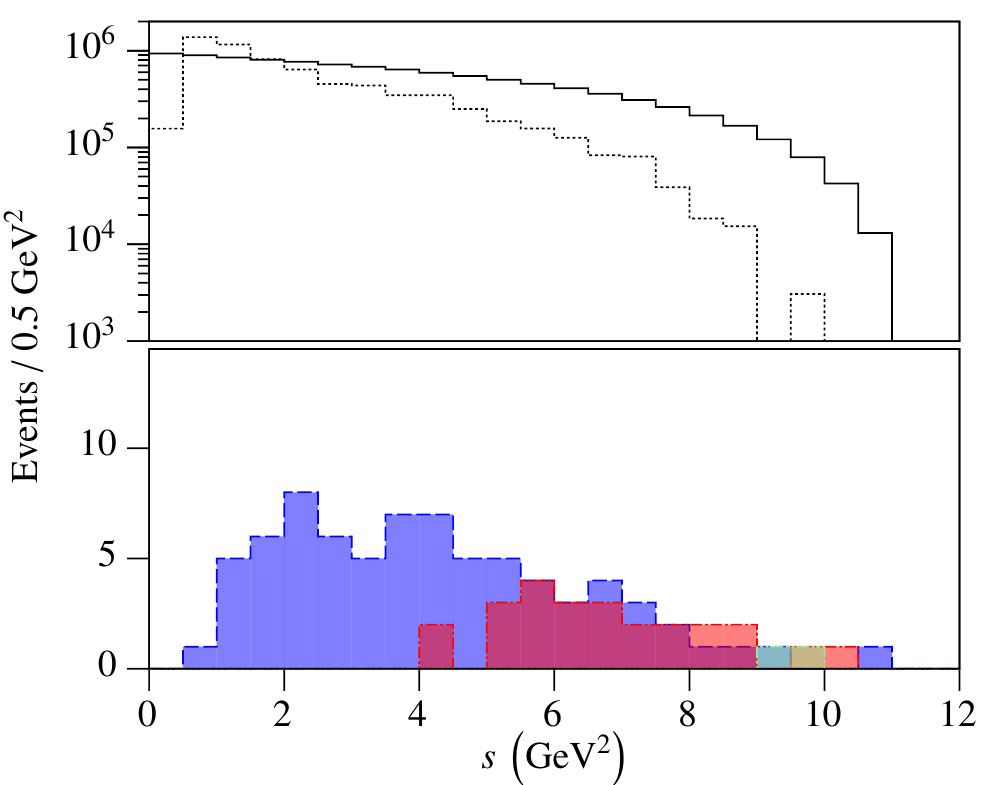}} \hfil%
\subfloat[Distribution of events with respect to $E_\mu$.
\label{fig:bg-Em}]{\includegraphics[width=0.45\linewidth,keepaspectratio]{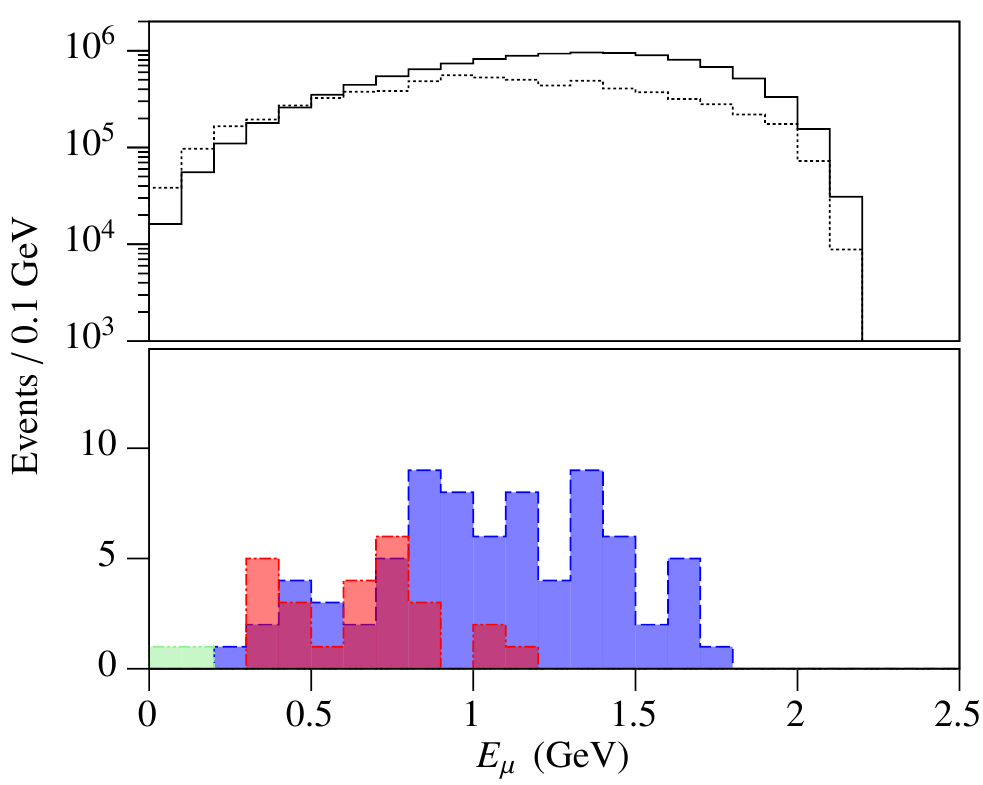}}\\%
\subfloat[Distribution of events with respect to $E_\textrm{miss}$.
\label{fig:bg-EN}]{\includegraphics[width=0.45\linewidth,keepaspectratio]{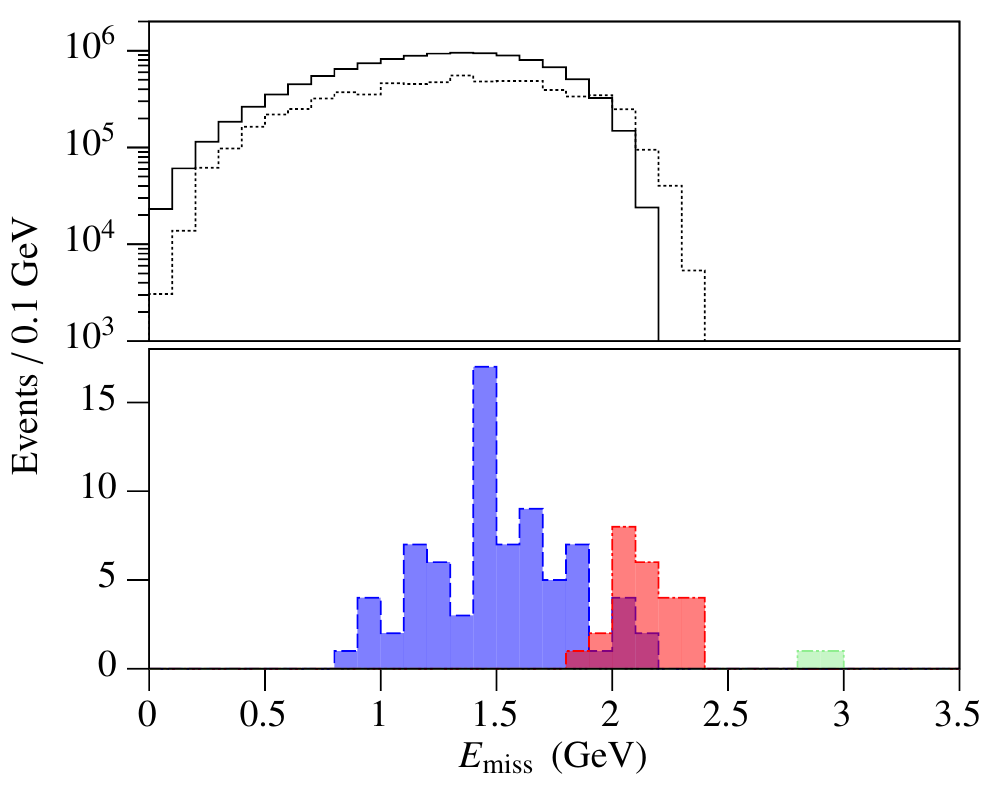}} \hfil%
\subfloat[Distribution of events with respect to $m_\textrm{miss}$.
\label{fig:bg-mN}]{\includegraphics[width=0.45\linewidth,keepaspectratio]{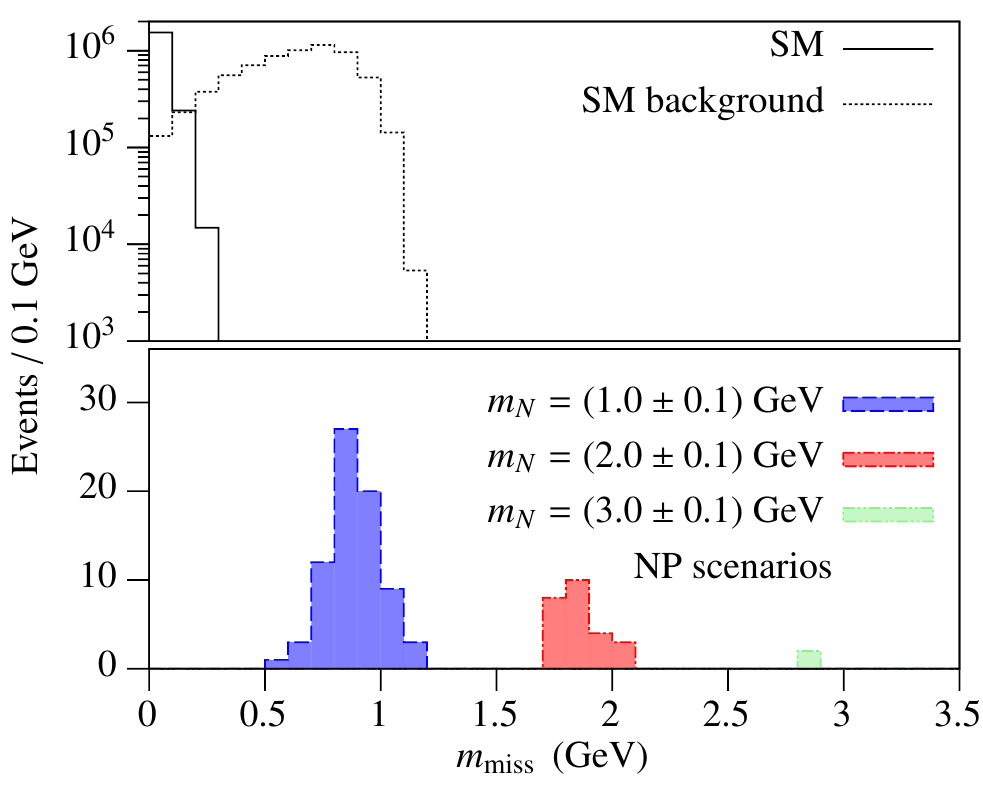}}%
\caption{Distribution of events corresponding to SM process $B \to D \mu
\nu_\mu$, SM background process $B \to D \mu \nu_\mu
\pi_\textrm{soft}/\gamma_\textrm{soft}$, and the new physics (NP) decays $B \to
D \mu N$ for $m_N = \left( 1.0, 2.0, 3.0 \right) \pm 0.1$~GeV,
with respect to a few observables. Note that the number of events for SM and SM
background processes are very large and, hence, those are shown with the
vertical axis in log-scale, while the NP scenarios are shown in a linear
scale.}%
\label{fig:background}
\end{figure*}

At this point, it is important to discuss how the signal events can be
experimentally distinguished from possible SM background events. To illustrate
our approach, we shall consider the signal processes $B \to D \mu N$ for $m_N =
\left( 1.0, 2.0, 3.0 \right) \pm 0.1$~GeV, the SM allowed process $B \to D \mu
\nu_\mu$, and the SM background process $B \to D \mu \nu_\mu \pi_\textrm{soft} /
\gamma_\textrm{soft}$, where we have taken pions with energy $< 0.2$~GeV and
photons with energy $< 0.1$~GeV in the rest frame of the $B$-meson as soft pions
and soft photons respectively. If we consider the existing experimental limits
on $\modulus{U_{\mu N}}^2$, then for the cases of $m_N=1.0,2.0,3.0$~GeV we
expect to obtain about $1,1,6$ signal events\footnote{Those numbers of signal
events can be easily seen from Fig.~\ref{fig:UmNSq}.}, respectively, which are
too few for any proper analysis. Besides, to keep our proposal for the new
experimental search devoid of any prejudice and yet optimistic, let us fix the
value of $\modulus{U_{\mu N}}^2$ at $10^{-5}$ for all values of $m_N$ under our
consideration. This yields about $75, 25, 2$ signal events for
$m_N=1.0,2.0,3.0$~GeV cases, respectively\footnote{See
Fig.~\ref{fig:background}, and please note that we have considered for each mass
$m_N$ the number of events$/0.1$ GeV.}. In Fig.~\ref{fig:background} we compare
the signal event distributions vs.\ the SM events and SM background events, with
respect to the energy of muon ($E_\mu$), the missing energy ($E_\textrm{miss}$,
with $E_\textrm{miss} = E_N$ for signal events), the missing mass
($m_\textrm{miss}$, with $m_\textrm{miss} = m_N$ for signal events) and the
invariant mass-square $s=\left(p_B - p_D\right)^2$. It is clear from
Fig.~\ref{fig:background} that the missing mass distribution is the most useful
one among all the observables, as the NP scenarios with $m_N > 1$~GeV
are easily discernible. Nevertheless, combinations of all the observables could
be used to look for the sterile neutrino signature. Since the SM and SM
background processes have much larger statistics and they are well understood
both theoretically and experimentally, one could, in principle, implement
multi-variate analysis or likelihood studies to figure out NP cases for $m_N <
1$~GeV.

We have demonstrated how the active-sterile mixing parameters can be probed
without considering any sequential decay of sterile neutrino, provided the
4-momenta of all the other particles are well measured. However, as mentioned
before and as is well known, the Dirac and Majorana nature of the sterile
neutrino can be probed only when its sequential decay inside detector is
considered.

\section{Probing the Dirac and Majorana nature of the neutrino $N$}\label{sec:majorana}

\begin{figure}[htb]
\centering%
\subfloat[Both Dirac and Majorana $N$]{\includegraphics[width=0.495\linewidth]{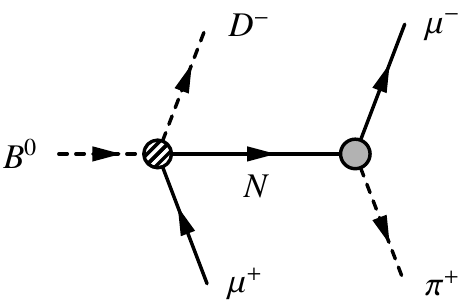}} \hfill%
\subfloat[Only Majorana $N$]{\includegraphics[width=0.4955\linewidth]{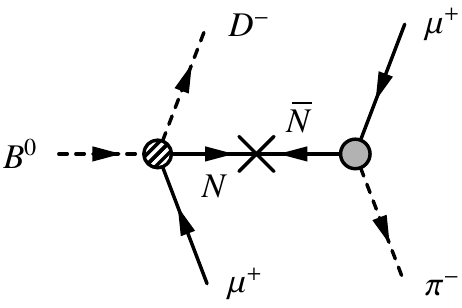}}%
\caption{Meson-level Feynman diagrams contributing to the decays $B^0 \to D^-
\mu^+ \mu^\mp \pi^\pm$. The sterile neutrino is produced at the first vertex and
decays at the second vertex, which is at an observable distance away from the
first vertex. The circular blobs connote the contributions from the
corresponding hadronic form factors and decay constants. The cross in the
Majorana scenario denotes the helicity flip involved in the decay.}%
\label{fig:Feynman-diagrams}
\end{figure}

If the sterile neutrino $N$ decays inside a detector, we can probe lepton number
violation in the entire process (which includes both the production of $N$ and
its sequential decay) to ascertain its Majorana nature. As an example, let us
consider the sequential decay, $B^0 \to D^- \mu^+ \mu^\mp \pi^\pm \equiv \left(
B^0 \to D^- \mu^+ N\right) \otimes \left(N \to \mu^\mp \pi^\pm \right)$. The
meson-level Feynman diagrams for these decays are shown in
Fig.~\ref{fig:Feynman-diagrams}. It is very clear that observation of the lepton
number violating mode $B^0 \to D^- \mu^+ \mu^+ \pi^-$ would imply that the
sterile neutrino has Majorana nature. While reconstructing the sterile neutrino
from the final states $\mu^{\mp} \pi^{\pm}$ in the detector, we must also
include an observable spatial separation between the point of production and
point of decay of the sterile neutrino. We can also consider the decay $N \to
\tau^\mp \pi^\pm$ if allowed by kinematics. Similar analysis as above can be
done for $D^0 \to K^- \mu^+ \mu^\mp \pi^\pm$ and related decays as well.

For a detector of finite size, say $L_D$, the observation of displaced vertex
with decay length $L$ necessarily demands that $L < L_D$, and this depends on
the lifetime and energy of the sterile neutrino.

The feasibility of studying the Dirac and Majorana signatures of the sterile
neutrino $N$ (of mass $m_N$, energy $E_N$ and total decay rate $\Gamma_N$) via
$B \to D \mu \mu \pi$ decays by using a detector of finite size $L_D$ depends on
two important factors, (1)~$P_\text{decay} (L)$, the probability of decay of $N$
within $L < L_D$, and (2)~$P_\text{flip}$, the probability of helicity flip
required for observation of the LNV decays which characterize the Majorana
neutrino, and these are given by
\begin{subequations}
\begin{align}
P_\text{decay} (m_N,E_N,L) &= 1 - \exp\left(- L \, m_N \, \Gamma_N/\sqrt{E_N^2 -
	m_N^2}\right),\\%
P_\text{\!flip} (m_N,E_N) &= m_N^2/\left( E_N + \sqrt{E_N^2 - m_N^2} \right)^2.
\end{align}
\end{subequations}
We can quantify the feasibility of observing the full decay $B \to D \mu \mu
\pi$ at Belle~II, by the following distribution of events with respect to $E_N$,
\begin{subequations}
\begin{align}
\frac{d\mathcal{N}_{LNC} \left(m_N,E_N,L\right)}{dE_N} &= \frac{N_B}{\Gamma_B} \frac{d\Gamma\left(B \to D
\mu N\right)}{dE_N} \times \Br\left(N \to \mu \pi\right) \nonumber\\%
& \quad \times P_\text{decay} (m_N,E_N,L) \nonumber\\%
& \quad \times \epsilon_D \times \epsilon_{\mu,1} \times \epsilon_{\mu,2} \times \epsilon_\pi,\\%
\frac{d\mathcal{N}_{LNV} \left(m_N,E_N,L\right)}{dE_N} &= \frac{d\mathcal{N}_{LNC} \left(m_N,E_N,L\right)}{dE_N} \times P_\text{flip} \left(m_N,E_N\right),
\end{align}
\end{subequations}
where $\mathcal{N}_{LNC}, \mathcal{N}_{LNV}$, respectively, denote the number of
LNC and LNV events, $N_B$ is the total number of $B$ mesons produced/analyzed in
the experiment, $\Gamma_B$ is the total decay rate of the $B$ meson, and
$\epsilon_D, \epsilon_{\mu,1},\epsilon_{\mu,2},\epsilon_\pi$ are the various
efficiency factors corresponding to the reconstruction of final $D$ meson, $\mu$
from the first vertex, $\mu$ from the second vertex and the $\pi$, respectively.
For our numerical study we have assumed all these efficiency factors to be $1$.
It should be noted that inclusion of $P_\text{\!flip} (m_N,E_N)$ is an effective
way of introducing the helicity flip. The most accurate way is to consider the
full decay $B \to D \mu \mu \pi$ along with the propagator for the intermediate
$N$ and that would automatically lead to $m_N$ dependence (similar to the
mass-dependence found in neutrinoless double-beta decay, a classic example of a
LNV process). By including helicity-flip factor in $d\mathcal{N}_{LNV}
\left(m_N,E_N,L\right)/dE_N$ it is clear that for $m_N \to 0$, the difference
between Dirac and Majorana cases vanish as per the `practical Dirac-Majorana
confusion theorem'.

\begin{figure}[hbtp]
\centering%
\includegraphics[width=0.9\linewidth,keepaspectratio]{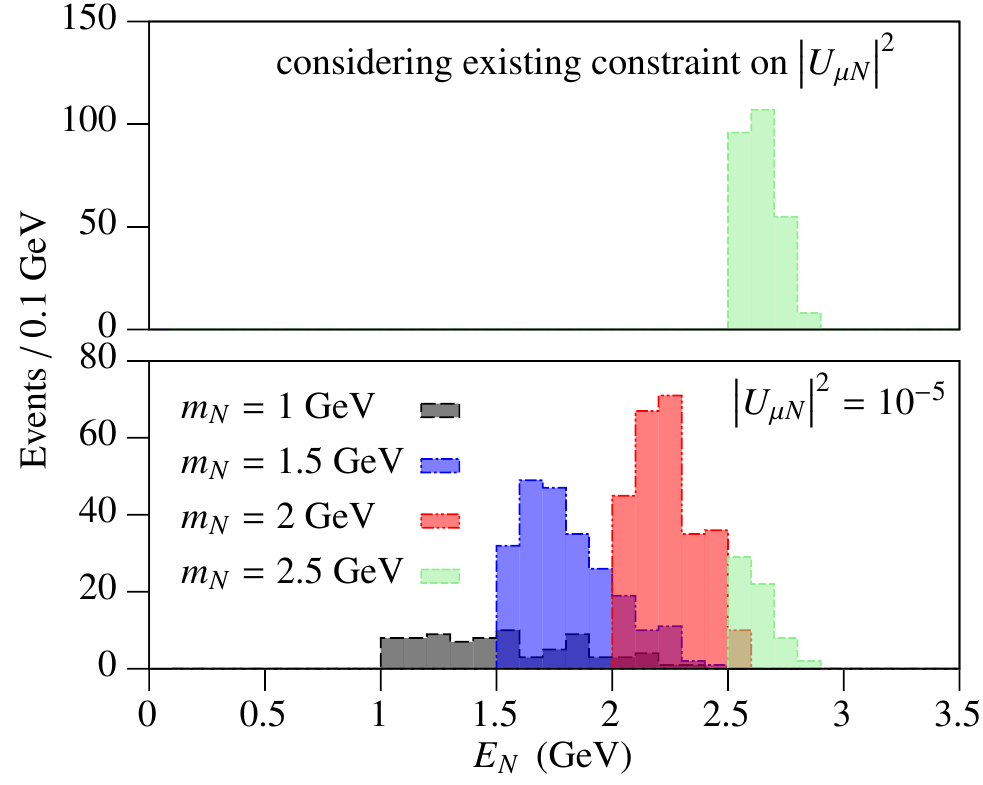}%
\caption{Distribution of number of events of the lepton number conserving $B \to
D \mu^\pm \mu^\mp \pi$ decays. Here we have considered the displaced vertices to
lie within $1$~m so that the events can be observed at Belle~II.}%
\label{fig:NumDensity-FullEvents}
\end{figure}

\begin{figure*}[h]
\centering%
\includegraphics[width=0.7\linewidth,keepaspectratio]{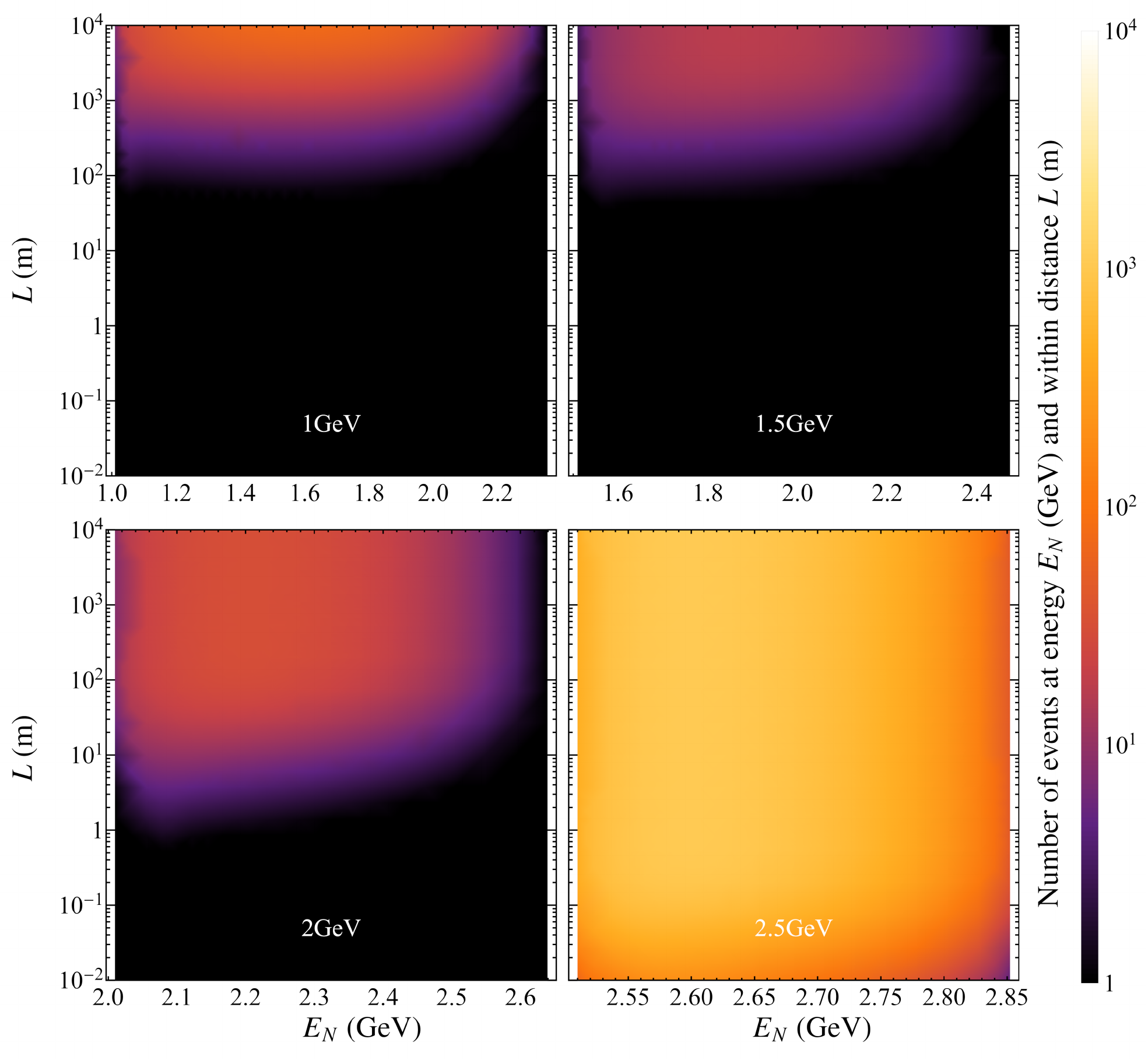}%
\caption{Distribution of events for various energies of the neutrino as measured
in the $B$ rest frame and considering the events with various displaced vertices
of lengths. Here the decays happen within the distance $L$ and only those decays
with $L \leqslant 1$~m are currently feasible for observation at Belle~II. We
have considered four benchmark cases corresponding to $m_N=1,1.5,2,2.5$~GeV. The
neutrino $N$ is assumed to have Dirac nature and the existing constraint on
$\modulus{U_{\mu N}}^2$ is considered here. The black colored region corresponds
to number of events $\leqslant 1$.}%
\label{fig:num-density-EN-L}%
\end{figure*}

\begin{figure*}[h]
\centering%
\includegraphics[scale=0.8]{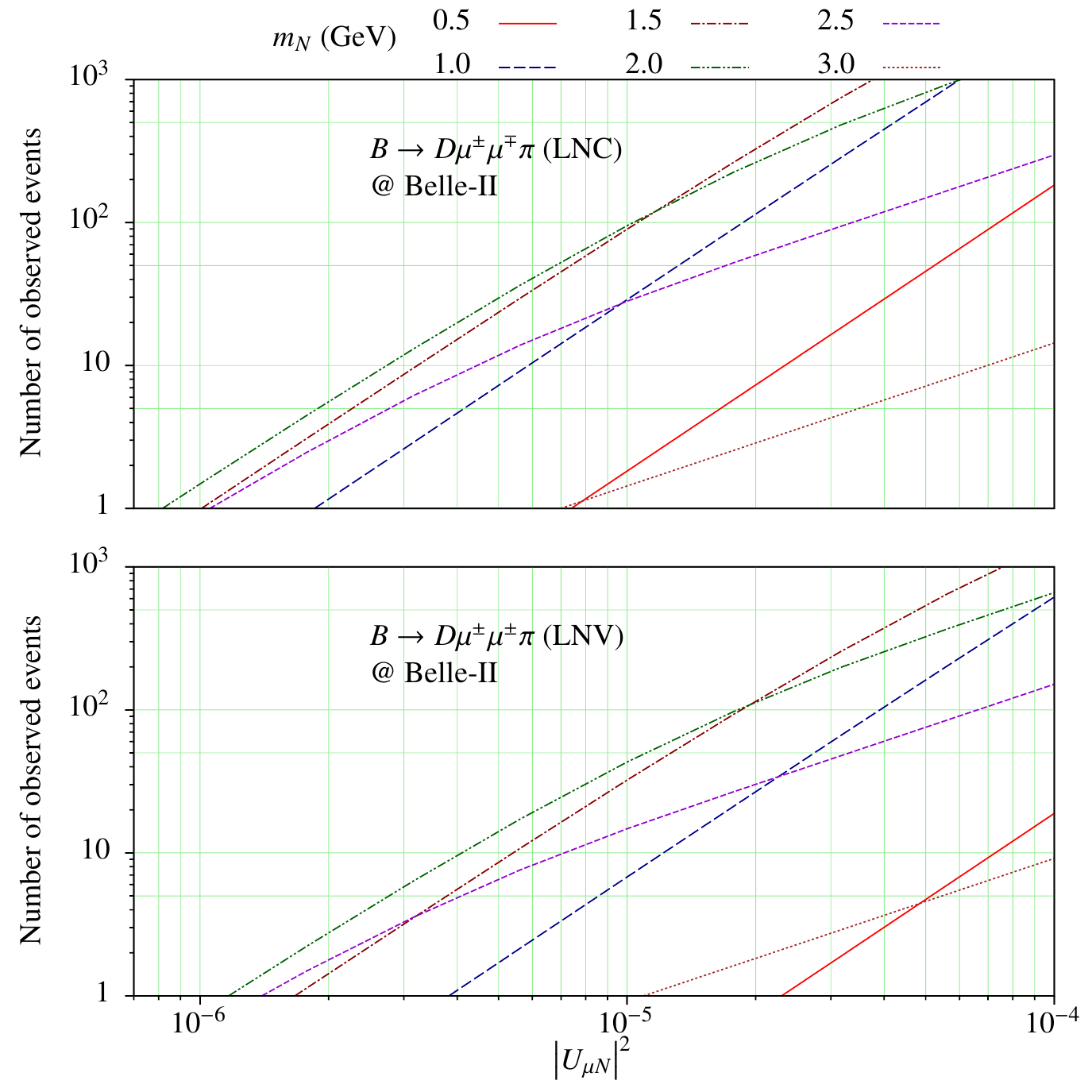}%
\caption{Numerical study of feasibility of observing purely Dirac (LNC) signal
$B \to D \mu^\pm \mu^\mp \pi$ and purely Majorana (LNV) signal $B \to D \mu^\pm
\mu^\pm \pi$, inside Belle~II detector with decay lengths less than $1$~m. Here
we have neglected the contributions from both $\modulus{U_{eN}}^2$ and
$\modulus{U_{\tau N}}^2$ when compared with $\modulus{U_{\mu N}}^2$.}%
\label{fig:feasibility}%
\end{figure*}

In the numerical study shown in Fig.~\ref{fig:NumDensity-FullEvents} we have
studied the energy distribution of the $\mu\pi$ system originating from the
decay of $N$. It is noticeable that for  $m_N \leqslant 2$~GeV scenarios, no $B
\to D \mu^\pm \mu^\mp \pi$ decays with characteristic displaced vertices
$\leqslant 1$~m are expected to be observable at Belle~II (considering the size
of the central drift chamber \cite{Kou:2018nap}) if we consider the existing
constraints on $\modulus{U_{\mu N}}^2$ (see
Refs.~\cite{Deppisch:2015qwa,Cvetic:2019shl} for details of existing
constraints). However, if we consider $\modulus{U_{\mu N}}^2 = 10^{-5}$, the
smaller mass scenarios are also feasible. It must be noted that the number of
events shown here are larger than the ones shown in Fig.~\ref{fig:bg-EN} simply
because we have considered the full sample of $B$ mesons here whereas for
Fig.~\ref{fig:bg-EN} only the smaller number of fully reconstructed $B$ decays
were considered. Fig.~\ref{fig:NumDensity-FullEvents} shows that as the mass of
neutrino gets smaller the mean distance of displaced vertices gets bigger which
would put most of the decays outside the Belle~II detector, unless there is
compensation by a substantial increase in the value of $\modulus{U_{\mu N}}^2$
which facilitates an appreciable number of events still happening inside the
detector. This is also clearly discernible from Fig.~\ref{fig:num-density-EN-L}
where we have varied the decay length $L$ within which the observed displaced
vertices would lie. As per the existing constraint on $\modulus{U_{\mu N}}^2$ if
we wish to observe $B \to D \mu^\mp \mu^\pm \pi$ events in a detector for $m_N
\leqslant 2$~GeV, our existing detectors are clearly not suitable. Nevertheless,
our approach elaborated in previous sections might come handy in the search for
discovery of such sterile neutrino(s). It should be noted that the energy $E_N$
in both Figs.~\ref{fig:NumDensity-FullEvents} and \ref{fig:num-density-EN-L} are
measured in the rest frame of the parent $B$ meson.
Fig.~\ref{fig:num-density-EN-L} \textcolor{black}{also illustrates an interesting
behavior that the probability of decaying within a smaller
decay length is larger when the neutrino is less energetic or equivalently more
non-relativistic. Moreover, as we go to higher masses, the neutrino decays faster
with larger decay width and shorter life time.}

Finally assuming that one can observe the full decays at Belle~II, we can
estimate the value of the active-sterile mixing parameter $\modulus{U_{\mu
N}}^2$ as a function of observed number of events. In Fig.~\ref{fig:feasibility}
we have analyzed how the number of events varies for different values of $m_N$
and $\modulus{U_{\mu N}}^2$. We have also considered the additional suppression
from helicity flip factor $P_\text{flip}$ while considering the LNV mode. For
the total decay rate of $N$ which enters $P_\text{decay}(L)$ we have used
Eqs.~(30--32) of Ref.~\cite{BtoDlN} and it depends on $\modulus{U_{eN}}^2$,
$\modulus{U_{\mu N}}^2$ and $\modulus{U_{\tau N}}^2$. Since $\modulus{U_{eN}}^2$
is already constrained by $0\nu\beta\beta$ experiments to be much smaller than
$\modulus{U_{\mu N}}^2$ and $\modulus{U_{\tau N}}^2$ (see
Ref.~\cite{Deppisch:2015qwa}), we can safely neglect its contribution. It can be
inferred from Fig.~\ref{fig:feasibility} that for $\modulus{U_{\mu N}}^2$
smaller than the existing experimental upper limit, one can still aspire to
observe more than a handful of $B \to D \mu \mu \pi$ decays at Belle~II with
displaced vertex signatures.

\section{Conclusion}\label{sec:conclusion}

In this paper we offer a new strategy that could lead to the discovery of a
heavy sterile neutrino $N$ of mass $m_N \leqslant 3.3$~GeV and having
appreciable mixing with active neutrinos, at the Belle~II experiment.
Consideration of the defining properties of a sterile neutrino such as it being
an electrically neutral, spin-$\nicefrac{1}{2}$ fermion with non-zero mass and
possibly long lifetime, irrespective of its Dirac or Majorana nature, points out
that decays such as $B\to D\ell N$, with $\ell=\mu,\tau$ and without the
sequential decay of $N$, could play a crucial role in the discovery of $N$. Once
$N$ is discovered, it is possible to probe the Majorana nature via searching for
the lepton number violating processes.

We have shown that at Belle~II, with only about $4.8 \times 10^8$ events of
fully reconstructed $B \to D \mu N$ decays, one can probe the mixing parameter
$\modulus{U_{\mu N}}^2$ to a precision which is comparable with what LHCb can
probe with about $4.8 \times 10^{12}$ events of $B \to D^{(*)}\mu \mu \pi$
decays. Our approach exploits the fact that at Belle~II the 4-momentum of the
parent $B$ meson decaying to $D \mu N$ can be inferred from the 4-momentum of
the accompanying $\overline{B}$ meson, a feat not feasible at LHCb. It is
noteworthy that for $m_N > 1.0$~GeV, it is easy to remove possible SM background
contamination. For $m_N < 1.0$~GeV, a more refined analysis of SM background
would be helpful in extracting the signal events.

It is certainly alluring to consider the full decay final states of $B \to D \mu
\mu \pi$ as they can be used to probe Dirac or Majorana nature of $N$. However,
these decays can have other new physics contributions in addition to
contribution from $N$. Considering the contribution from $N$ as the major one,
such decays still suffer suppression from displaced vertices that must lie
within the finite sized detectors, as well as from branching ratio of the
sequential $N$ decay and possible spin-flip associated with the Majorana
signature. Thus, despite their undisputed strength, these suppressed decays
would be difficult to observe and discovery of $N$ might be missed if such
decays are considered. For some masses, especially the heavier mass of $N$,
however, we can still expect to observe a few such events even at Belle~II,
albeit the lower number of $B$ mesons produced as compared to LHCb.

In summary, in this paper we propose the most systematic approach to probe a
sterile neutrino $N$ of mass $m_N \leqslant 3.3$ GeV, by using the decays $B\to
D\ell N$ (with $\ell=\mu,\tau$) rather than the suppressed $B \to \ell_1 \ell_2
\pi, X \ell_1 \ell_2 \pi$ decays (with $\ell_{1,2}=e,\mu$ and $X=\pi,D$)
previously used at Belle and LHCb\cite{BtoXll}. Our proposal does not require
the sequential decay of $N$. The constraint on $\lvert U_{\mu N} \rvert^2$ thus
achievable from Belle~II is not only better than the existing experimental
constraints in certain mass range of $m_N$ (viz.\ $0.4$--$1$~GeV and
$2$--$3$~GeV), but also better than the constraint that is achievable at
upgraded LHCb for $m_N < 2$~GeV. The minimum value of $m_N$ that can be probed
is constrained only by experimental accuracy of measurement of 4-momenta of $B$,
$D$ and $\mu$. The sequential decay of $N$, useful to distinguish its Dirac or
Majorana nature by observing the LNC or LNV modes respectively, is suppressed
from observation of displaced vertices (for both LNC and LNV cases) and helicity
flip (for LNV case only). Our numerical study shows if no decays of $N$ get
observed within decay lengths~$\leqslant 1$~m, the existing experimental upper
limit on $\lvert U_{\mu N} \rvert^2$ can be improved.

\begin{acknowledgement}
The work of C.S.K. was supported in part by the National Research Foundation of
Korea (NRF) grant funded by the Korean government (MSIP) (NRF2018R1A4A1025334).
This work of D.S. was supported (in part) by the Yonsei University Research Fund
(Post Doc. Researcher Supporting Program) of 2018 (project no.: 2018-12-0145).
\end{acknowledgement}


\end{document}